\documentclass[a4paper,11pt]{article}
\usepackage{jinstpub} 
\usepackage{lineno}

\title{\boldmath Development and extension of a monochromatic neutron beamline for neutron polarimetry device characterization at the Spallation Neutron Source}

 \author[a,b,1]{K. Imam,\note{Corresponding author.}}
 \author[c]{V. Cianciolo,}
 \author[d]{B. Filippone,}
 \author[b]{N. Fomin,}
 \author[b,c]{G. Greene,}
 \author[c]{C. Jiang,}
 \author[b]{J. O'Kronley,}
 \author[c]{S. Penttila,}
 \author[c]{J. Pierce,} 
 \author[c]{J. Ramsey}
 \author[b]{and I. Wallace}
 \affiliation[a]{Lawrence Livermore National Laboratory,\\Livermore, California, USA}
 \affiliation[b]{University of Tennessee,\\Knoxville, Tennessee, USA}
 \affiliation[c]{Oak Ridge National Laboratory,\\Oak Ridge, Tennessee, USA}
 \affiliation[d]{Kellogg Radiation Laboratory, California Institute of Technology,\\Pasadena, California, USA}

\emailAdd{imam2@llnl.gov}

\abstract{The precise manipulation and analysis of neutron spin states are foundational for a wide range of physics experiments, from fundamental symmetry tests to materials science. To enable systematic characterization of neutron polarimetry devices, we have extended an existing monochromatic neutron beamline at the Spallation Neutron Source (SNS), Oak Ridge National Laboratory. The beamline delivers monochromatic neutrons and provides a flexible platform for deploying and evaluating advanced neutron spin manipulation instruments. We describe the design and commissioning of the extended beamline, and present a proof-of-concept neutron polarimetry study using three neutron polarimetry devices: the supermirror neutron polarizer, the Mezei spin flipper, and an in situ polarized $^{3}$He-based neutron spin analyzer system. Performance metrics, optimization strategies, and systematic effects are discussed, demonstrating the beamline’s utility for neutron instrumentation testing. Our results establish the extended monochromatic beamline as a useful resource for the development and validation of neutron polarimetry technologies.}


\arxivnumber{2602.00167} 

\begin{document}
\maketitle
\flushbottom

\section{Introduction}
\label{sec:intro}

Neutron beams with well-defined spin states are essential for probing magnetic phenomena, understanding fundamental symmetries, and advancing neutron scattering techniques. Searches for the neutron's electric dipole moment (nEDM) require detailed understanding and control of magnetic fields to minimize systematic effects; polarized neutron transport serves as a sensitive probe for characterizing these fields, and the performance of polarimetry devices for this characterization therefore impacts the sensitivity of such measurements to new sources of fundamental symmetry violation~\cite{Ahmed2019}. Polarized neutron beta decay experiments, such as the pNab experiment, rely on precise knowledge of the beam polarization, measured using polarimetry devices, to extract fundamental weak interaction parameters~\cite{Baessler2025}. In condensed matter physics, polarized neutron scattering techniques depend on high-performance polarization hardware for resolving magnetic structures and spin excitations in complex materials~\cite{Wang2018}. The ability to produce, manipulate, and analyze polarized neutron beams relies on a suite of sophisticated polarimetry devices such as spin filters, polarizers, and spin flippers. The performance and reliability of these devices must be rigorously characterized under realistic beam conditions to ensure their suitability for deployment in high precision experiments. 

Dedicated beamlines for the development and testing of neutron polarization instrumentation exist at neutron science facilities such as the HB-2D Polarized Neutron Development beamline~\cite{Crow2016} at the High Flux Isotope Reactor (HFIR) and PHADES (Polarized $^{3}$He And Detector Experiment Station)~\cite{nist_phades} at the NIST Center for Neutron Research. These facilities have demonstrated the value of dedicated beamlines for neutron polarization R\&D. In this light, a complementary capability at the Spallation Neutron Source (SNS) is desirable for the following reasons: the pulsed time-of-flight beam structure of the SNS differs fundamentally from reactor-based continuous sources, and devices intended for deployment on SNS instruments benefit from characterization under the same beam conditions they will encounter in production. A dedicated polarimetry test beamline at the SNS therefore fills an important gap, enabling device optimization and cross-calibration in a pulsed spallation source environment.

At the SNS, a monochromatic neutron beamline has been constructed to provide wavelength selected neutrons for a variety of applications. Recognizing the need for a dedicated beamline to test and optimize neutron polarimetry devices, we have extended this beamline, creating a flexible experimental area where devices can be installed, aligned, and systematically studied. This extension includes a neutron flight tube, collimation and shielding, and a suite of diagnostic detectors, enabling precise control over beam properties and environmental conditions.

In this paper, we describe the design and extension of the monochromatic neutron beamline, and present a proof-of-concept polarimetry measurement using three neutron polarimetry devices: the supermirror neutron polarizer, the Mezei spin flipper, and an in situ polarized $^{3}$He-based neutron spin analyzer. We detail the principles of operation, experimental configurations, and results for each device, and discuss the implications for neutron instrumentation development. Our work demonstrates the value of a dedicated monochromatic beamline for advancing neutron polarimetry and supports the broader goals of fundamental neutron physics and materials research.




\section{Monochromatic Neutron Beamline}
\label{sec:monobeam}

The SNS monochromatic neutron beamline is engineered to primarily deliver 8.9~\AA\ neutrons, selected via alkali-intercalated graphite monochromators and filtered by a time-of-flight frame overlap chopper~\cite{Fomin2015}. The beamline receives cold neutrons from a liquid hydrogen moderator, with a time averaged proton power of up to 1.6 MW~\cite{Fomin2015}. Downstream of the monochromators, the beam is guided through an expanding half-ellipsoid neutron guide, emerging into a shielded experimental enclosure~\cite{Fomin2015}. This upstream monochromatic beamline and its 8.9~\AA\ neutron production capability were established previously~\cite{Fomin2015}.

\begin{figure}[htbp]
\centering
\includegraphics[width=\textwidth]{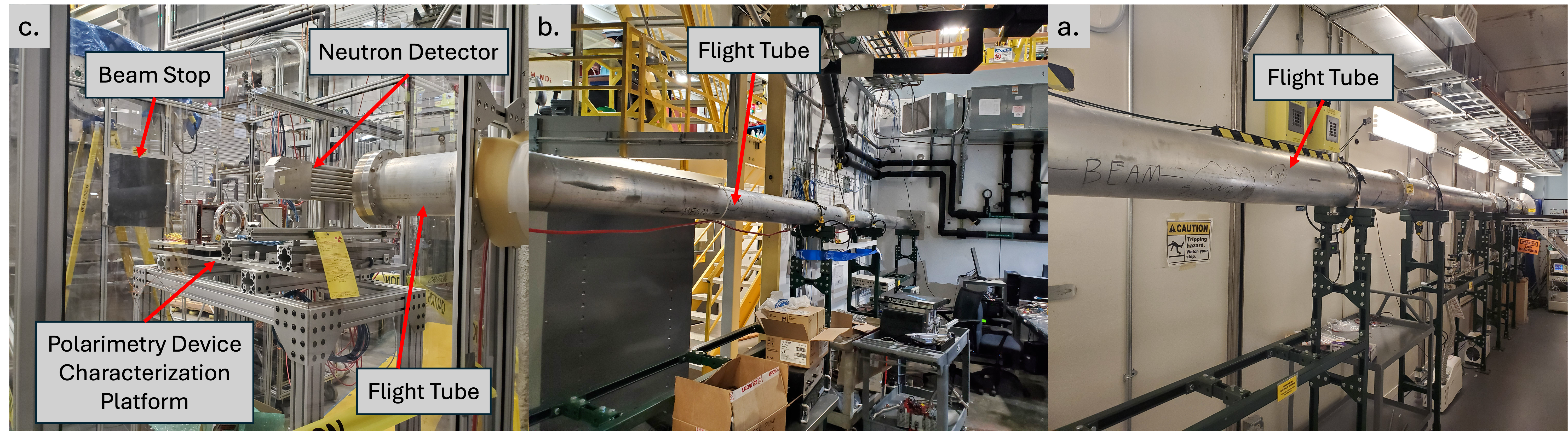}
\caption{The extended monochromatic neutron beamline at the SNS. (a) The flight tube originates from the expanding half-ellipsoid neutron guide in the shielded experimental enclosure. (b) The extension consists of an aluminum flight tube with lithium apertures and borated shielding, connecting the enclosure to the external experimental hall where neutron polarimetry devices are installed. (c) The terminus of the flight tube and the device characterization platform, where polarimetry devices can be deployed for testing and characterization.}
\label{fig:beamline_schematic}
\end{figure}

To facilitate neutron polarimetry studies, we have extended the beamline with a flight tube constructed from aluminum pipe. The flight tube, as shown in figure~\ref{fig:beamline_schematic}, connects the shielded enclosure to an external experimental hall, providing a controlled environment for neutron polarization device installation and beam manipulation. Collimation is achieved using lithium apertures and skimmers, which define the beam profile and minimize activation and background. Borated shielding lines the flight tube to suppress stray neutrons and ensure radiation safety.

\begin{figure}[htbp]
\centering
\includegraphics[width=1.1\textwidth]{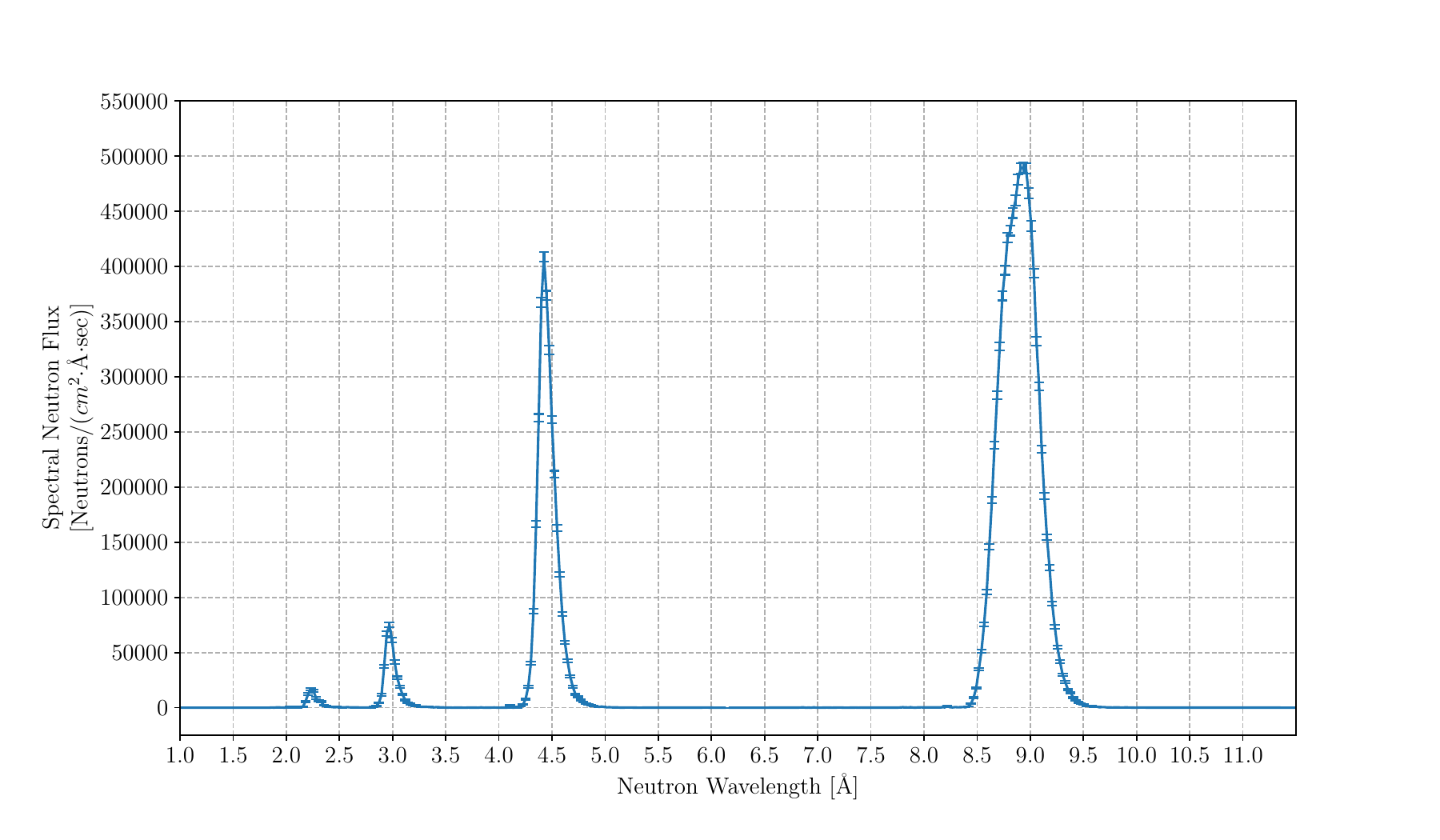}
\caption{Spectrum from the SNS monochromatic neutron beamline measured as a function of neutron wavelength.}
\label{fig:13Aspectrum}
\end{figure}

Beam spectrum characterization was performed using a calibrated $^{3}$He proportional counter while neutron counting for beam polarization measurements was performed using a position sensitive 8-pack $^{3}$He tube detector array~\cite{Berry2012, Diawara2023}. Time-of-flight (TOF) measurements, synchronized with the SNS accelerator’s proton pulse timing, enable precise determination of neutron wavelength and flux. The SNS-wide EPICS-based data acquisition system (DAQ) was used to collect the time of flight event mode data from the detectors~\cite{Vodopivec2018}. The DAQ also provides the live proton power from the SNS accelerator to normalize the flux, accounting for any beam drop. The neutron spectrum from the monochromatic beamline measured during the January 2022 SNS beam production cycle is shown in figure~\ref{fig:13Aspectrum}. The figure shows the primary 8.9~\AA\ neutron beam with the secondary peaks ($\lambda/2$, $\lambda/3$ and $\lambda/4$) reflected by the monochromator as well. The time-of-flight frame overlap chopper was used to filter out the secondary neutron wavelengths. The extended beamline supports flexible experimental geometries, allowing for upstream and downstream placement of polarimetry devices, detectors, and auxiliary equipment.

\section{Neutron Polarimetry Device Characterization}
\label{sec:devchar}

For the proof-of-concept polarimetry measurements, the extended monochromatic beamline supports two complementary purposes: individual characterization of neutron polarimetry devices and integrated beam-polarization measurements. In this section, we first summarize the operating principles and standalone performance of the in situ polarized 
$^{3}$He-based neutron spin analyzer, the supermirror polarizer, and the Mezei spin flipper using the 8.9~\AA\ monochromatic neutron beam. Section \ref{sec:fullpol} then combines these devices in the full beamline configuration to measure the neutron beam polarization. The measurements reported here were performed during the August 2023 SNS beam production cycle.

\subsection{In Situ Neutron Spin Analyzer System}

Neutron transmission measurements through the in situ polarized $^{3}$He-based neutron spin analyzer (polarized and unpolarized configurations) enable determination of neutron polarization. A key feature of the $^{3}$He analyzer system is its strong spin-dependent neutron absorption, which enables it to function as an efficient neutron spin filter (NSF)~\cite{Chupp1987, Greene1995, Musgrave2018}. The capture reaction $^{3}$He(n, p)t has a very large cross section for neutrons with spins antiparallel to the $^{3}$He polarization, and is highly suppressed for parallel spins. This strong spin selectivity allows a polarized $^{3}$He cell to transmit one neutron spin state while absorbing the other, making it a useful tool for neutron polarimetry.

\begin{figure}[htbp]
\centering
\includegraphics[width=\textwidth]{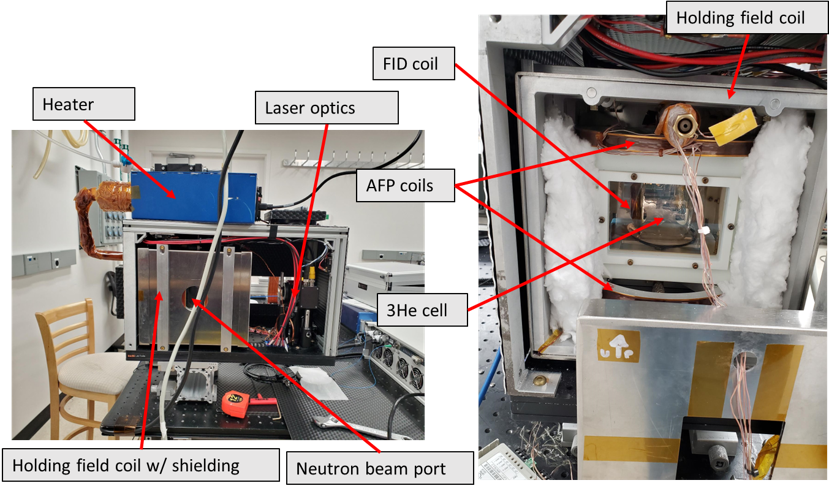}
\caption{The in situ polarized $^3$He neutron spin analyzer system, showing the SEOP laser, the $^3$He cell within the Merritt four-coil configuration and mu-metal shielding, and the fused silica windows for simultaneous optical pumping and neutron transmission. The system also contains NMR coils for monitoring and manipulating $^3$He polarization.}
\label{fig:analyzer_photo}
\end{figure}

The in situ neutron spin analyzer system was developed at ORNL for live neutron spin filtering and to replace the less operationally convenient drop-in polarized $^{3}$He spin analyzer method~\cite{Jiang2014, Jiang2017, Jiang2023}. The in situ neutron spin analyzer system is designed to maintain a continuously polarized $^{3}$He filled cell directly in the neutron beam path. The in situ neutron spin analyzer system used for the measurements reported in this study is shown in figure~\ref{fig:analyzer_photo}.

To generate polarized $^{3}$He, the system is integrated with a Spin Exchange Optical Pumping (SEOP) setup. The setup includes a circularly polarized laser (50 W, 770 nm) which illuminates the cell containing small amounts of alkali mixture (Rb/K) in addition to $^{3}$He. The cell, which is made of aluminosilicate glass, is heated to vaporize the alkali mixture (Rb/K), and a uniform holding magnetic field (12.7 G) is generated using a Merritt four-coil configuration and mu-metal shielding. The laser optically pumps the alkali mixture, which transfers its angular momentum to $^{3}$He. The in situ system has fused silica windows, allowing for simultaneous neutron beam transmission for spin analysis as well as continuous SEOP operation to maintain $^{3}$He polarization. 

Free-induction decay (FID) and adiabatic fast passage (AFP) diagnostics on the cell enable real-time monitoring and manipulation of $^{3}$He polarization, ensuring stable operation and facilitating calibration. During operation, FID signals exhibited long $T_{2}^{*}$ times, indicating low magnetic field gradients and stable $^{3}$He polarization. AFP sweeps performed to invert the $^{3}$He polarization also indicated minimal loss (0.053\% per flip), enabling repeated spin analysis.

The analyzing power of the $^{3}$He cell, which quantifies its ability to discriminate neutron spin, was first determined using the transmission of neutron beam through an unpolarized and polarized $^{3}$He cell using the scheme provided in refs.~\cite{Chupp1987, Greene1995}. In our beamline tests, the $^{3}$He cell was characterized in situ by measuring the transmission of 8.9~\AA\ neutrons through both unpolarized and polarized cell states. The measurements are listed in table~\ref{tab:3HeAnalPower}. For the cell used, with 0.8 bar pressure and 6.62 cm length, the analyzing power, $P_n^{^{3}He}$, and the absolute $^{3}$He polarization, $P_{^{3}He}$, were measured as:
\begin{align} 
P_n^{^{3}He} &= 0.70 \pm 0.07_{stat} \\ 
P_{^{3}He} &= 0.27\pm 0.04_{stat}
\end{align}
respectively, relative to unity. These results confirm the strong spin selectivity of the analyzer and provide a direct calibration for neutron polarization measurements on the testbed beamline.

\begin{table}[htbp]
\centering
\caption{Summary of the transmission measurements of 8.9~\AA\ unpolarized neutron beam through the different configurations of polarized $^{3}$He cell. Here, $N_0$ denotes the open-beam count rate with no analyzer present, $T_{unpol}$ denotes the open-beam count rate with unpolarized $^{3}$He cell and $T_{pol}$ denotes the open-beam count rate with polarized $^{3}$He cell.}
\smallskip
\begin{tabular}{@{}lccc@{}}
\hline
State & Beam Polarization State & Analyzer Polarization State & Neutrons/(s $\cdot$ MW) \\
\hline
$N_0$ & Unpolarized & Not Present & 4489 \\
$T_{unpol}$ & Unpolarized & Unpolarized & 189 \\
$T_{pol}$ & Unpolarized & Polarized & 263 \\ 
\hline
\end{tabular}
\label{tab:3HeAnalPower}
\end{table}

\subsection{Supermirror Polarizer}

For polarizing the neutron beam, the currently existing supermirror neutron polarizer at the fundamental neutron physics beamline was utilized~\cite{Balascuta2012}. The supermirror neutron polarizer is a multilayer device engineered to reflect and transmit neutrons based on their spin state. It consists of alternating layers of ferromagnetic iron (Fe) and nonmagnetic silicon (Si) deposited on borofloat glass panes. The multilayer structure creates a spin-dependent optical potential, with a critical angle for total reflection that differs for spin-up and spin-down neutrons. The ``m'' factor of the supermirror (here, m=3) denotes the enhancement of the critical reflection angle relative to a pure $^{58}$Ni surface. This allows efficient polarization of neutrons over a wide range of incident angles and wavelengths. The magnetic field from permanent magnets around the polarizer saturates the magnetization of Fe layers, establishing a uniform magnetic field and maximizing spin selectivity.

In the beamline extension, the polarizer is mounted on a rotation and translation stage, allowing fine adjustment of its yaw, pitch, and position relative to the beam axis. The device comprises 45 curved borofloat panes, each 0.3 mm thick, with a 9.6 m radius of curvature~\cite{Balascuta2012}. This geometry ensures that all neutrons experience at least one bounce, eliminating line-of-sight transmission and maximizing polarization efficiency. Alignment is achieved by tuning the yaw angle for maximum transmission and polarization.

\subsection{Mezei Spin Flipper}

\begin{figure}[htbp]
\centering
\includegraphics[width=0.75\textwidth]{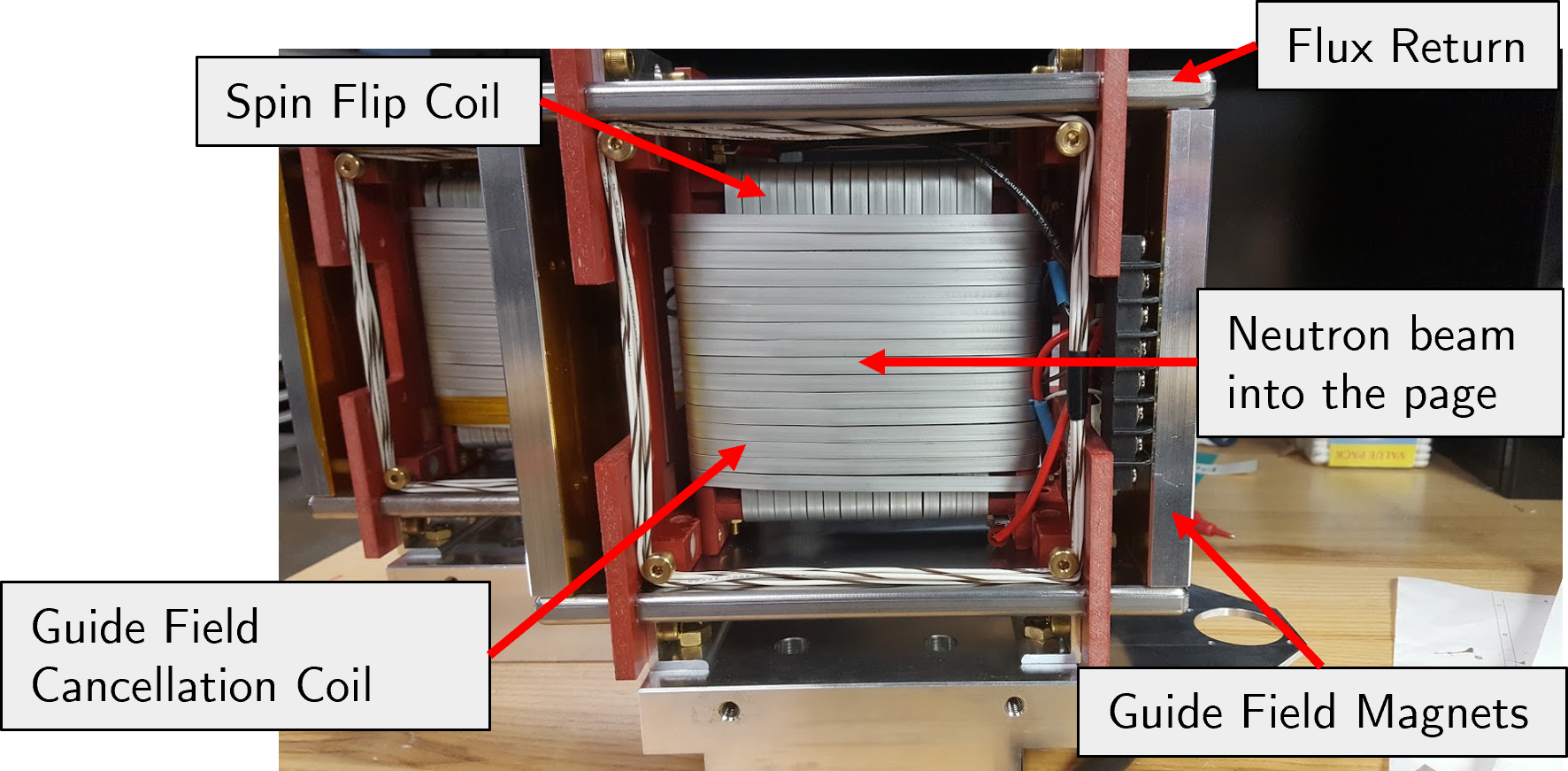}
\caption{The Mezei spin flipper showing the rectangular aluminum ribbon solenoid coil, the outer guide field cancellation coil, and the orthogonal shim coils for fine field adjustment.}
\label{fig:flipper_photo}
\end{figure}

The Mezei spin flipper is a diabatic device designed to invert the spin state of polarized neutrons by applying a strong transverse magnetic field relative to the guide field~\cite{Mezei1972}. The flipper consists of a rectangular solenoid coil (constructed from aluminum ribbon wire), an outer coil for guide field cancellation, permanent magnets for field return, and orthogonal shim coils for fine adjustment. When powered, the solenoid coil generates a magnetic field perpendicular to the neutron beam’s guide field. As neutrons traverse the flipper, the transverse field induces a torque on their magnetic moment, causing a spin rotation of $n\pi$ radians, where $n$ is an odd integer. The flipping efficiency depends on the field strength, neutron velocity, and coil geometry. The Mezei spin flipper used for this characterization is shown in
figure~\ref{fig:flipper_photo}.

\begin{figure}[htbp]
\centering
\includegraphics[width=\textwidth]{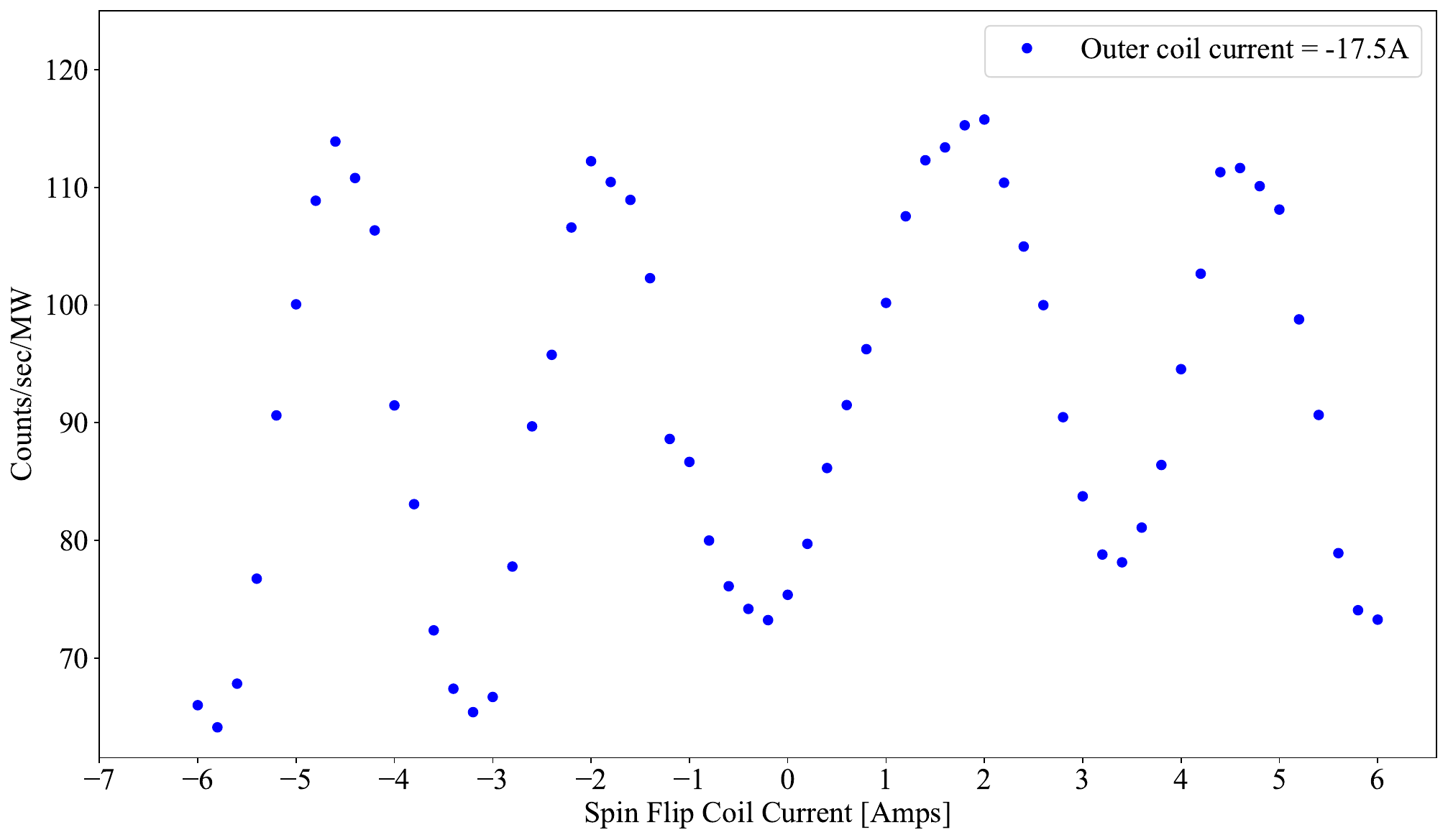}
\caption{8.9~\AA\ neutron count rate while sweeping the flipper coil current for the guide field cancellation coil current of -17.5 A.}
\label{fig:SF_test}
\end{figure}

The Mezei flipper utilized for this polarimetry measurement was installed downstream of the supermirror neutron polarizer, with its axis aligned perpendicular to the beam direction. The flipping coil current was adjusted, allowing for systematic sweeps to identify the optimal flipping ratio. The outer coil cancels the guide field, ensuring a purely transverse field during the flip. Performance was evaluated by measuring the transmission of neutrons through a downstream spin analyzer (e.g., $^{3}$He cell) as a function of flipping coil current. The flipping ratio (high/low transmission) and flipping efficiency are extracted from these measurements. Figure~\ref{fig:SF_test} shows the transmitted neutron count rate as a function of the solenoid coil current for a guide-field cancellation current of -17.5 A. The oscillatory dependence reflects the varying spin rotation angle produced by the flipper as the transverse field is changed. An operating point was selected to maximize spin contrast for 8.9~\AA\ neutrons. From these scans, we obtained a flipping efficiency of 0.93 and a flipping ratio greater than 1.7, demonstrating robust spin flipping as well as the ability to perform multiple spin turns under the beamline operating conditions.

\section{Neutron Beam Polarization Measurements}
\label{sec:fullpol}

\begin{figure}[htbp]
\centering
\includegraphics[width=\textwidth]{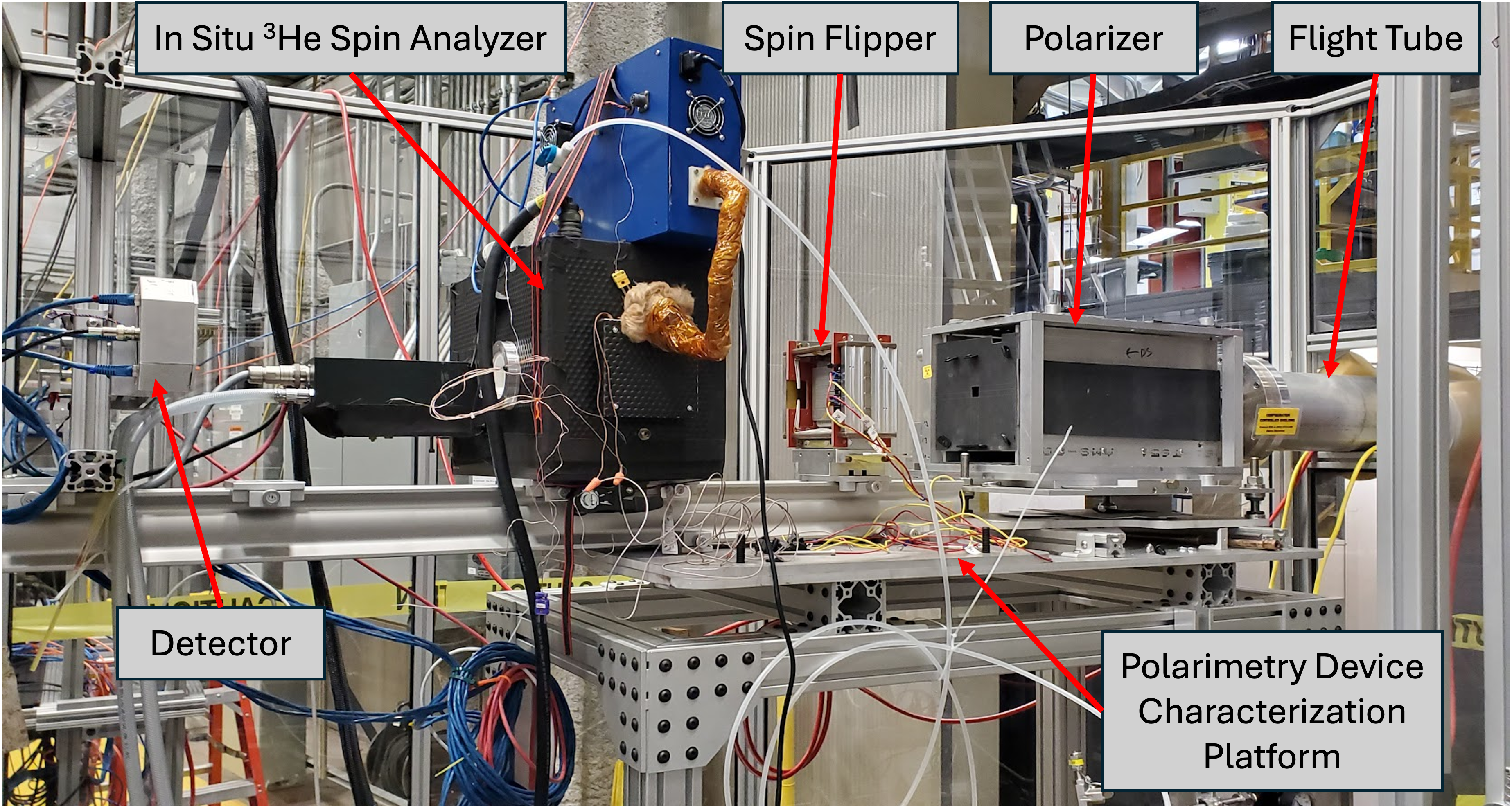}
\caption{Neutron polarimetry devices deployed on the extended monochromatic beamline. The incoming unpolarized neutron beam comes from the end of the flight tube and enters the supermirror polarizer, shown mounted on its rotation and translation stage. The neutron beam polarization gets analyzed via the in situ polarized $^3$He neutron spin analyzer system. The Mezei spin flipper can also be activated to analyze both spin states of the neutron beam. An 8-pack $^3$He tube detector was used to measure the neutron counts. All components were mounted on a platform using rail mounts on an optical rail on-axis with neutron beam, allowing for positional and alignment adjustability.}
\label{fig:devices}
\end{figure}

A neutron beam polarization measurement was performed using the methodology outlined in reference~\cite{Musgrave2018}. The methodology involves measuring the transmission of a polarized neutron beam through an unpolarized, polarized, and AFP-reversed polarized $^{3}$He cell. The transmission of spin flipped neutrons through polarized and AFP-reversed polarized $^{3}$He cell is also measured to verify the spin contrast of the neutron beam. To do this, the extended monochromatic beamline was fitted with the supermirror polarizer, Mezei spin flipper, and the in situ spin analyzer, positioned in that order from upstream to downstream. The devices work in concert to produce, manipulate, and analyze polarized neutron beams with high efficiency and reliability. The supermirror polarizer generates a highly polarized beam. The Mezei flipper efficiently inverts the neutron spin, enabling measurement of both spin states and correction for systematic effects. The in situ neutron spin analyzer provides stable and continuous spin filtering, with integrated diagnostics for real-time calibration. A calibrated 8-pack $^{3}$He tube detector was used for neutron counting to measure the neutron transmission for each configuration. Figure~\ref{fig:devices} shows the experimental setup of the three neutron polarimetry devices used in the proof of concept neutron beam polarization measurement. The transmission count rates from the configurations are summarized in table~\ref{tab:polResult}. The neutron beam polarization, $P_{n}$, relative to unity, was measured as:
\begin{equation}
   P_{n}=0.83 \pm 0.07_{stat} 
\end{equation}

\begin{table}[htbp]
\centering
\caption{Summary of the transmission measurements of 8.9~\AA\ polarized neutron beam through the different configurations of polarized $^{3}$He cell. Here, $N_0$ denotes the open-beam count rate with no analyzer present, $T_0$ denotes transmission through the unpolarized $^{3}$He cell, $T$ denotes transmission through the polarized analyzer, $T_{sf}$ denotes transmission after neutron spin flip, and the superscript $AFP$ denotes reversal of the analyzer polarization by adiabatic fast passage.}
\smallskip
\begin{tabular}{@{}lccc@{}}
\hline
State & Beam Polarization State & Analyzer Polarization State & Neutrons/(s $\cdot$ MW) \\
\hline
$N_0$ & $\uparrow$ & Not Present & 36.8 $\pm$ 0.6 \\
$T_0$ & $\uparrow$ & Unpolarized &  1.85 $\pm$ 0.04 \\
$T$ & $\uparrow$ & $\uparrow$ &  3.57 $\pm$ 0.07 \\ 
$T_{sf}$ & $\downarrow$ & $\uparrow$ & 1.34 $\pm$ 0.03 \\
$T_{sf}^{AFP}$ & $\downarrow$ & $\downarrow$ & 3.25 $\pm$ 0.06 \\
$T^{AFP}$ & $\uparrow$ & $\downarrow$ & 1.21 $\pm$ 0.03 \\
\hline
\end{tabular}
\label{tab:polResult}
\end{table}

These measurements demonstrate that the extended monochromatic beamline can support both quantitative neutron polarimetry and systematic characterization of polarization hardware. In the present configuration, the in situ polarized $^{3}$He-based neutron spin analyzer provided strong spin selectivity, the Mezei flipper achieved a robust flipping efficiency, and the combined system yielded a neutron beam polarization for monochromatic neutrons. Future work will focus on improving magnetic-field uniformity, guide-field control, and magnetic field shielding in order to reduce systematic uncertainty and further increase the achievable beam polarization. Nevertheless, this was a key demonstration that neutron polarimetry and device characterization are feasible on the extended monochromatic beamline.

\section{Conclusion}
\label{sec:conc}

In this work, we have demonstrated the integration and characterization of a suite of neutron spin control and analysis devices—including a supermirror polarizer, Mezei spin flipper, and an in situ polarized $^{3}$He analyzer on a dedicated monochromatic neutron beamline. The combined system enabled control and measurement of neutron polarization, and provided a robust platform for evaluating device performance and identifying systematic effects.

The extended monochromatic beamline serves as a useful resource for the development and validation of neutron polarimetry technologies. Its flexibility and diagnostic capabilities enable systematic evaluation of device performance under realistic beam conditions. In fact, the extended beamline has already been used for the characterization of the adiabatic fast passage (AFP) spin flipper for the Nab experiment~\cite{Godri2025}. Lessons learned from these studies can be used to inform the design and deployment of polarimetry hardware in high-precision experiments, including neutron electric dipole moment searches, polarized neutron beta decay and polarized neutron scattering, supporting the broader goals of fundamental neutron physics and materials research.

\acknowledgments

The authors would like to thank K. Berry, L. Funk, A. Parizzi, X. Geng, B. Vacaliuc, M. Frost and E. Iverson of the Oak Ridge National Laboratory for providing technical support for this work. The work described in this article was in part supported by U.S. Department of Energy Office of Nuclear Physics award number DE-FG02-03ER41258, Office of Basic Energy Sciences, Early Career Research Program Award KC0402010, under Contract DE-AC05-00OR22725, Office of Science WDTS-SCGSR program under contract number DE‐SC0014664 and National Science Foundation award number NSF-2110898. This work was performed under the auspices of the U.S. Department of Energy by Lawrence Livermore National Laboratory under Contract DE-AC52-07NA27344. This manuscript	has been authored by Lawrence Livermore	National Security, LLC under Contract No. DE-AC52-07NA27344 with the US. Department of Energy. The United	States Government retains, and the publisher, by accepting the article for publication, acknowledges that the United States Government retains a non-exclusive, paid-up, irrevocable, world-wide license to publish or reproduce the published form of this manuscript, or allow others to do so, for United States Government purposes. LLNL-JRNL-2014898. 



\bibliographystyle{JHEP}
\bibliography{biblio.bib}






\end{document}